\documentclass[11pt]{article}
\usepackage{latexsym}
\usepackage{graphicx}
\usepackage{amsfonts,amssymb}

\begin{document}

\title{Empirical Analysis and Evolving Model of Bipartite Networks}
\author{Peng Zhang$^{1,2}$, Menghui Li$^{1}$, J.F.F. Mendes$^{2}$, Zengru Di$^{1}$, Ying Fan$^{1}$\footnote{Author for correspondence: yfan@bnu.edu.cn}, \\
\\ 1. Department of Systems Science, School of Management,\\
Beijing Normal University, Beijing 100875, P.R.China\\
2. Departamento de F\'{i}sica da Universidade de Aveiro,\\
3810-193 Aveiro, Portugal}

\maketitle

\begin{abstract}
Many real-world networks display a natural bipartite structure.
Investigating it based on the original structure is helpful to get
deep understanding about the networks. In this paper, some
real-world bipartite networks are collected and divided into two
types, dependence bipartite networks and independence bipartite
networks, according to the different relation of two sets of
nodes. By analyzing them, the results show that the actors nodes
have scale-free property in the dependence networks, and there is
no accordant degree distribution in the independence networks for
both two types of nodes. In order to understand the scale-free
property of actors in dependence networks, two growing bipartite
models without the preferential attachment principle are proposed.
The models show the scale-free phenomena in actors' degree
distribution. It also gives well qualitatively consistent behavior
with the empirical results.
\end{abstract}


\section{Introduction}\label{introduction}

In recent years, the complex networks has attracted more and more
people's attention \cite{Review1,SIAM,sergev,Albert}. Many
real-world systems are depicted as complex networks to investigate
their structures and functions. Examples include WWW, internet,
food webs, biochemical networks, social networks, and so on
\cite{sergev1,PNAS,jazz,Williams,emaila,emailb,overlapping}. These
researches in networks not only raised new concepts and methods,
but also helped us understand complex systems.

Bipartite network is an important kind of complex networks. In
fact, many real-world networks display a natural bipartite
structure, Examples are the actors-films network \cite{movie}, the
papers-scientists network \cite{paper,paper1,paper2} and so on. A
bipartite graph is defined as a network in which nodes are divided
into two sets, one representing a set of actors, while the other
representing acts of collaboration, so that no edge connects two
nodes in the same set. In other words, The edges only connect a
pair of nodes belongs to different sets. These networks are also
called 2-mode networks, or affiliation networks when they
represent groups and members ($i.e$. each link represents a social
actor's affiliation to a group) \cite{basic1,basic}. In the
present paper, following the relation of two sets of nodes, we
consider that bipartite networks can be separated into two types
of networks, dependence bipartite networks and independence
bipartite networks. In dependence networks, two sets are
dependent. The groups are the aftermaths of the behaviors among
actors, the actors lead to the groups. Examples of the dependency
networks are the actors-films network \cite{movie}, where a film
is a production of the cooperation among actors, the
papers-scientists network \cite{paper,paper1,paper2}, where some
scientists coauthor a paper, and the occurrence networks, where a
sentence is organized by some words. The independence network is
opposite with the dependence network, two sets exist
independently. Let us cite for instance the booker-reader network,
where the books always are as the resource in library whatever the
readers do, the sex networks \cite{sex}, where the male and the
female do not affect each other, and the peer to peer networks
\cite{p2p,p2p1}, where as the downloader and the uploader, they
are not the output of each other.

Now many real world networks have been analyzed as one mode
networks, and been found that have the scale-free property
\cite{paper1,paper2,self}. However most of them has not been done
on the generation of complex networks with the scale-free property
expressed by bipartite graphs. Newman $et\ al$ \cite{random}. have
proposed a model for generating a complex bipartite graph, in
which two degree distributions for both kinds of nodes are given
and edges are randomly connected under the constraint that their
degree distributions are fixed. Ramasco $et\ al$.\cite{self}
Guillaume $et\ al$\cite{model,model1}. Ramezanpour \cite{rame}.
Goldstein $et\ al$\cite{paper}. and Lambiotte $et\
al$\cite{lambiotte} have proposed the other type of models for
generation of complex bipartite graphs; those are growing networks
in which the number of nodes and that of edges are increasing and
newly added edges are connected to nodes by a suitable rule. In
order to generate a complex network of bipartite graph with the
scale-free property, Goldstein $et\ al$. Ramasco $et\ al$. and
Guillaume $et\ al$. have applied a rule, which is based on the
concept of the preferential attachment proposed by Barab\'{a}si
and Albert \cite{scale}, to their model for generating complex
networks. There also have the other type of bipartite networks in
which the size of networks is fixed, but the structure of networks
changes rapidly. Ohkubo $et\ al$ \cite{rewire}. note this and
proposed a model to generate complex bipartite graphs without
growing. The essential ingredients are a preferential rewiring
process and a fitness distribution function in this model. They
showed that the model generates the complex networks with the
scale-free property for some fitness distribution functions.

In the present paper, we collect different types of real world
data and build them to complex bipartite networks. By analyzing
these bipartite networks, we try to find the common properties in
the dependency networks (in the independence networks). Because in
bipartite networks, there is a lack of standard definitions of
some properties that we often analyze in one-mode networks, here
we only focus on the degree distribution of both types of nodes.
The results reveal that the actors nodes have scale-free property
in the dependence networks, and there is no accordant degree
distribution in the independence networks for both two types of
nodes. Then we propose two growing models of complex bipartite
networks for the dependence networks without preferential
attachment. The degree distribution of actors in the networks
obtained with both models can capture the scale-free property
which appears in real dependence networks.

The outline of this paper is as follow. In Section \ref{data}, we
introduce two types of networks in complex bipartite networks, the
dependence networks and the independence networks. Some real world
networks of them are built and analyzed. The results show that the
actors nodes have scale-free property in the dependence networks.
We propose two growing models of complex bipartite networks for
the dependence networks in Section \ref{model}. It also contains
the simulation results. We obtain the degree distribution of
generated dependence networks, and the scale-free property is
found in them. In Section \ref{conclusion}, we give our concluding
remarks.

\section{Bipartite Networks Data}\label{data}
In order to investigate the properties of the dependence networks
and the independence networks, here we consider some
representation real-world examples of them. The set of the
dependence networks includes a Econophysicists bipartite network
\cite{economy}, a complex-networks network, the co-occurrence
relation of words in the sentences of the Bible network
\cite{word}, and a coauthoring relation between scientists network
\cite{paper1,paper2,random,basic}. The independence networks
consist of a books-readers database obtained from Beijing Normal
University library, a P2P network gotten from $Maze.com$, and
another P2P network \footnote{The choice of two kind nodes are
different in these two P2P networks, and the data also gotten from
different sites.}\cite{p2p,p2p1}. To be convenient in next
sections, we refer to replace them as $Econophysicists$,
$Complex-networks$, $Cooccurrence$, $Coauthoring$, $Library$,
$Maze$, and $P2P$. All of them have been defined and studied as
the one-mode network well in cited references.

\subsection{The Dependence Networks}

{\bf{Econophysicists}}\ \  We collected publish information of
Econophysicists papers published from $1992$ to $2007$, and built
the Econophysicists bipartite network with if the author wrote the
paper, there would be an edge connects with them. The final
bipartite network comprises $N_{authors}=1963$ authors,
$N_{papers}=2020$ papers and $E=5129$ edges between them. The
average degree of authors and papers are
$<k_{authors}>=E/N_{authors}=2.61$ papers per author and
$<k_{papers}>=E/N_{papers}=2.54$ authors per paper, while the
maximal degree of them respectively are $85$ and $8$.

{\bf{Complex-networks}}\ \   The complex networks is a hot topic
in recent years. There are many papers about it. We search the
papers in SCI web sites with the keyword complex networks from
$1998$ to $2007$ to build this Complex-networks network.
Scientists and papers are identified as two different kinds of
nodes, an edge exists between a scientist and a paper if the
scientist is one of the paper's authors. the resulting bipartite
network consists of $50545$ scientists, $18833$ papers, and
$72582$ edges. The average degree of them are
$<k_{scientists}>=1.436$ and $<k_{papers}>=3.85$, while the
maximal degree are $83$ and $166$.

{\bf{Cooccurrence}}\ \  The data of this co-occurrence network
obtained from a new online version of the Bible. There are two
different kinds of nodes, words and sentences, with a word
connects to a sentence which it belongs to. The built bipartite
network is composed of $N_{words}=9264$, $N_{sentences}=13587$,
and $E=183363$. The average degree of words and sentences
respectively are $<k_{words}>=19.8$ and $<k_{sentences}>=13.5$,
while the maximal degree are $10454$ and $52$.

{\bf{Coauthoring}}\ \ This co-authoring bipartite network is
defined the same with the complex-networks network. The data
obtained from the online $arXiv$ preprint repository. It includes
$N_{scientists}=16400$, $N_{papers}=19885$, and $E=45904$. The
average degree of scientists and papers respectively are
$<k_{scientists}>=2.8$ and $<k_{papers}>=2.3$, while the maximal
degree are $62$ and $8$.

\begin{figure}
\includegraphics[width=7cm]{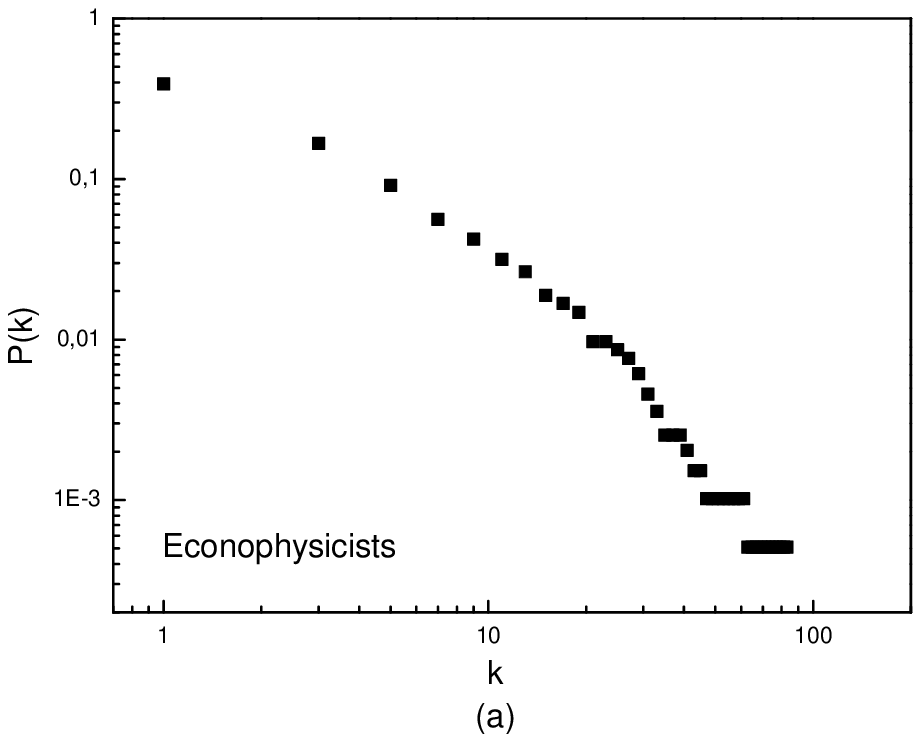}\includegraphics[width=7cm]{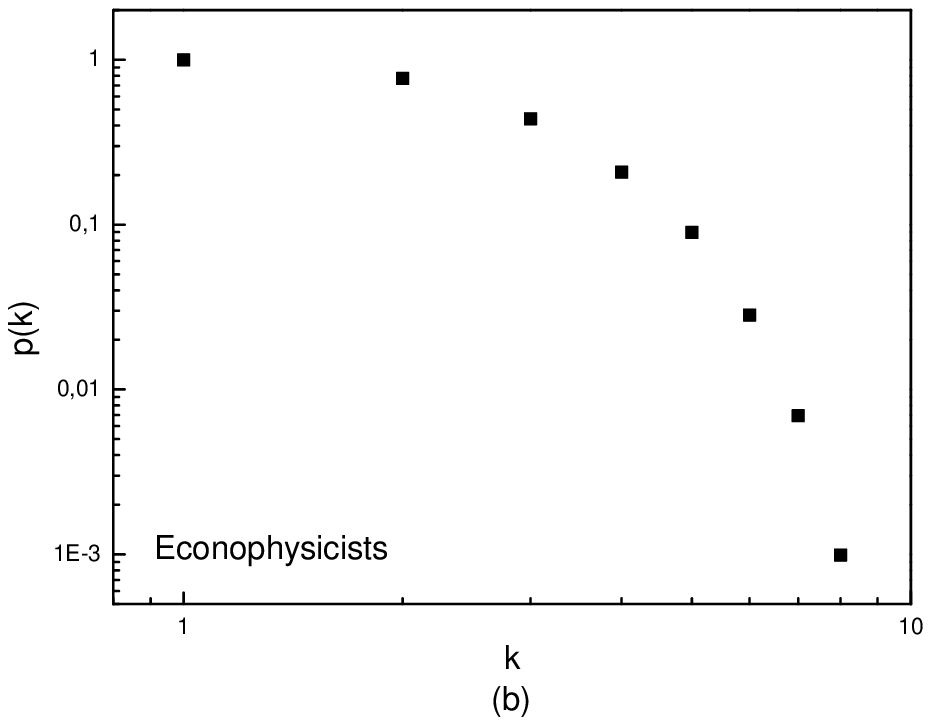}
\caption{(a) Degree distribution of authors in the Econophysicists
network. The degree $k$ corresponds to the number of papers of each
author signed. (b) The same distribution of papers in the
Econophysicists network. The degree $k$ corresponds to the number of
authors of each paper.}\label{Econophysicists}
\end{figure}

\begin{figure}
\includegraphics[width=7cm]{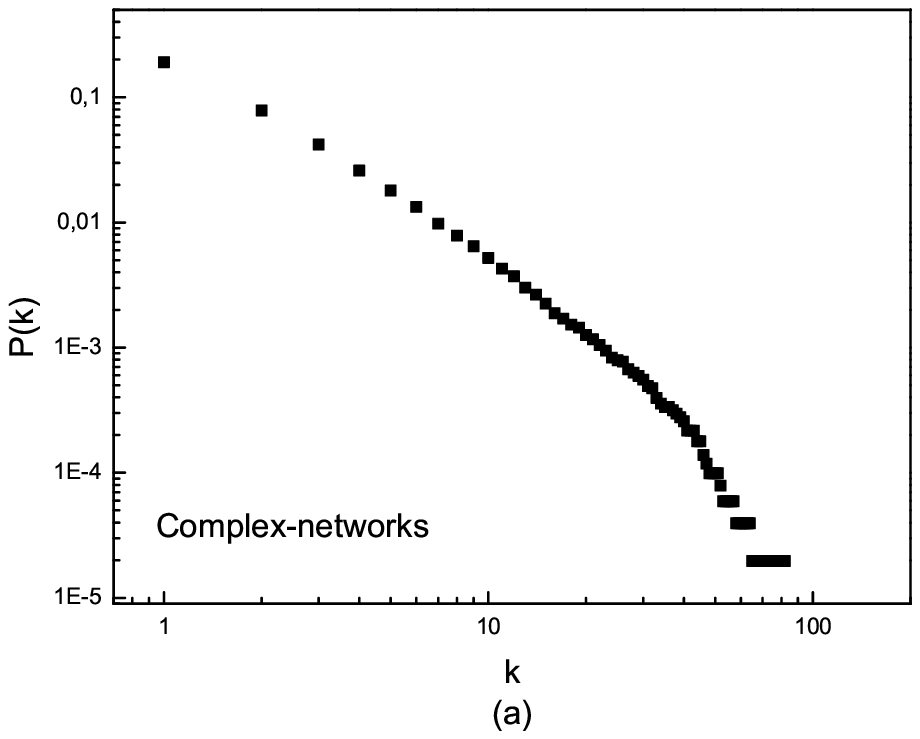}\includegraphics[width=7cm]{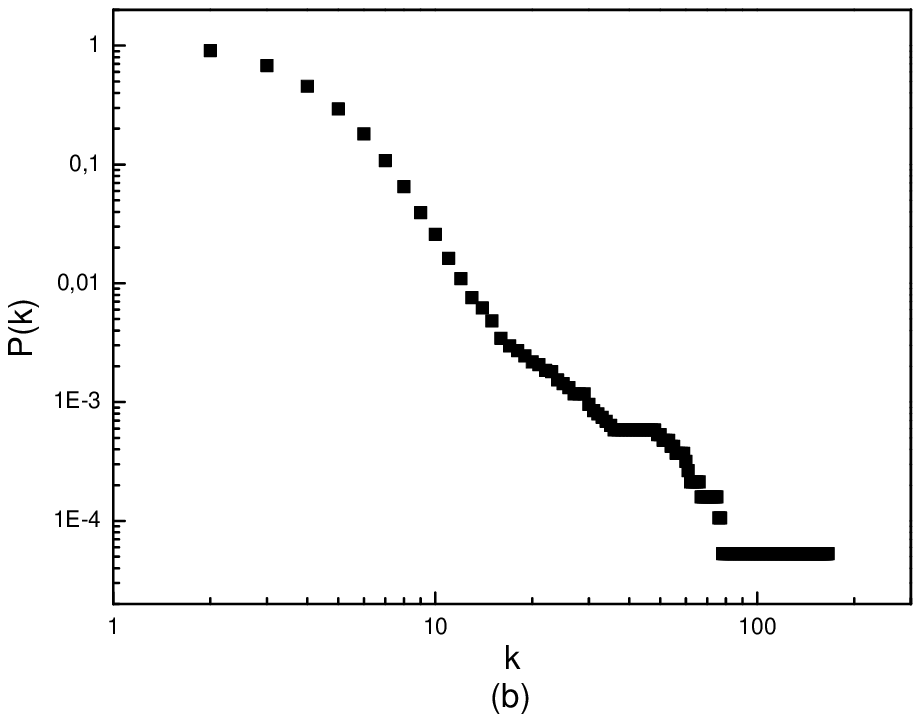}
\caption{(a) Degree distribution of scientists in the
Complex-networks network. The degree $k$ corresponds to the number
of papers of each scientist signed. (b) The same distribution of
papers in the Complex-networks network. The degree $k$ corresponds
to the number of scientists of each paper.}\label{Complex-networks}
\end{figure}

\begin{figure}
\includegraphics[width=7cm]{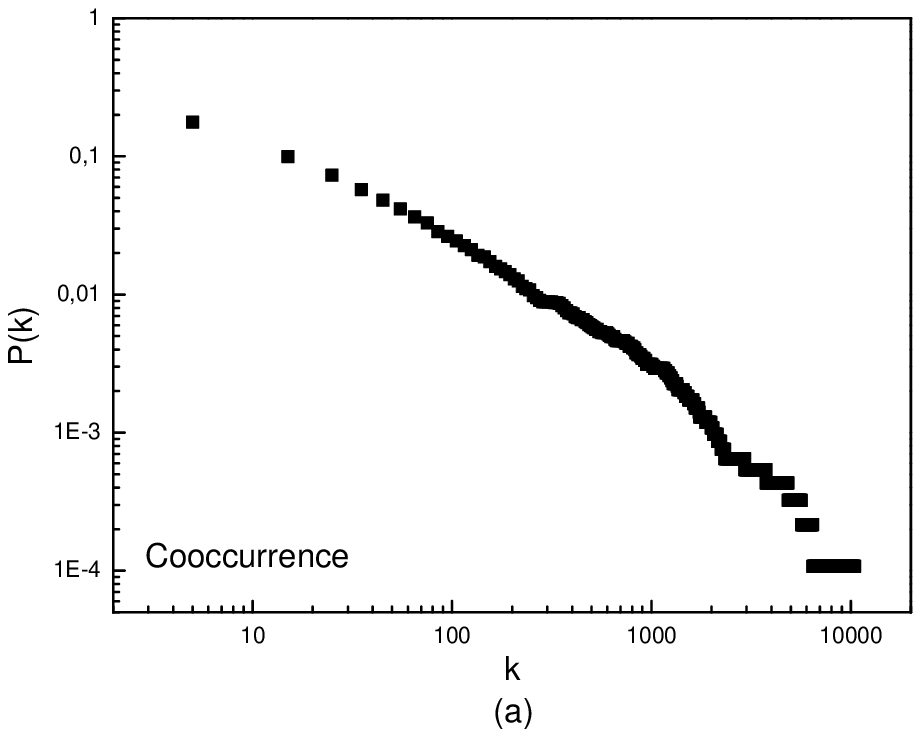}\includegraphics[width=7cm]{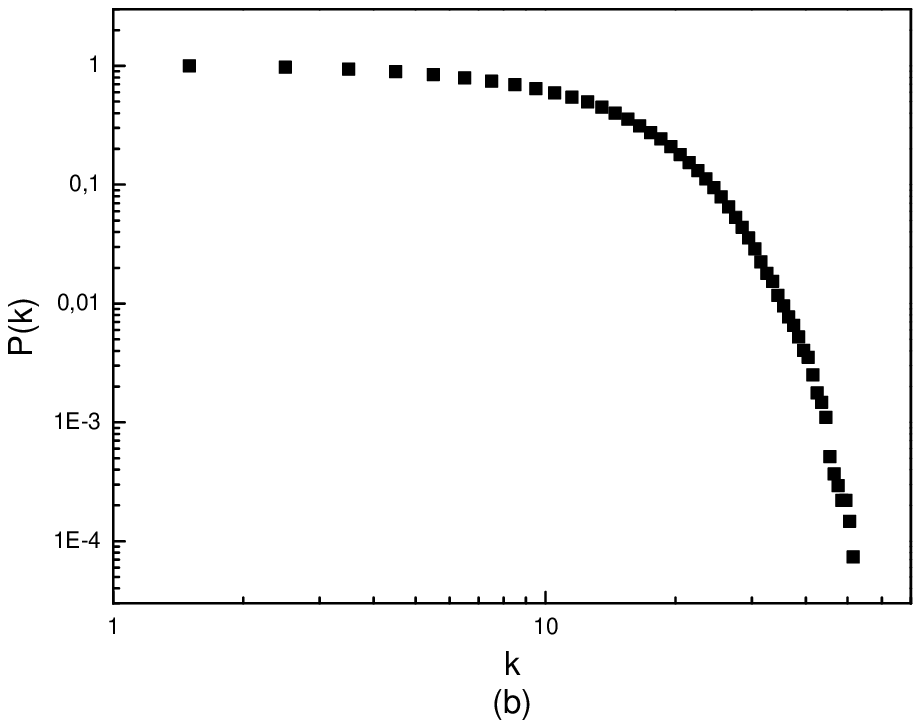}
\caption{(a) Degree distribution of words in the Cooccurrence
network. The degree $k$ corresponds to the number of sentences of
each word belonged to. (b) The same distribution of sentences in the
Cooccurrence network. The degree $k$ corresponds to the number of
words of each sentence.}\label{Cooccurrence}
\end{figure}

\begin{figure}
\includegraphics[width=7cm]{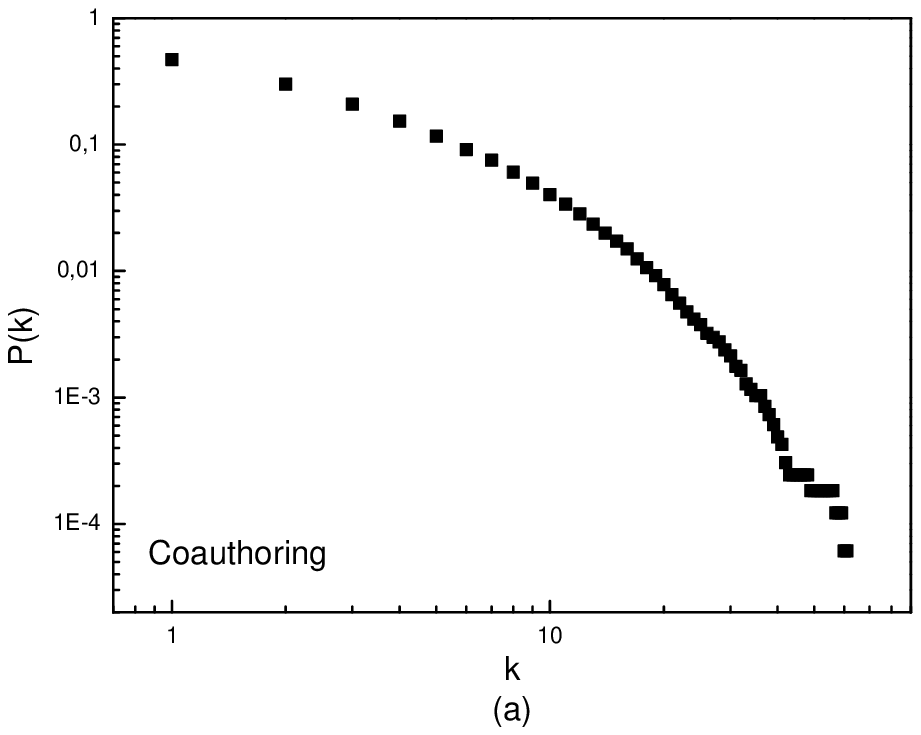}\includegraphics[width=7cm]{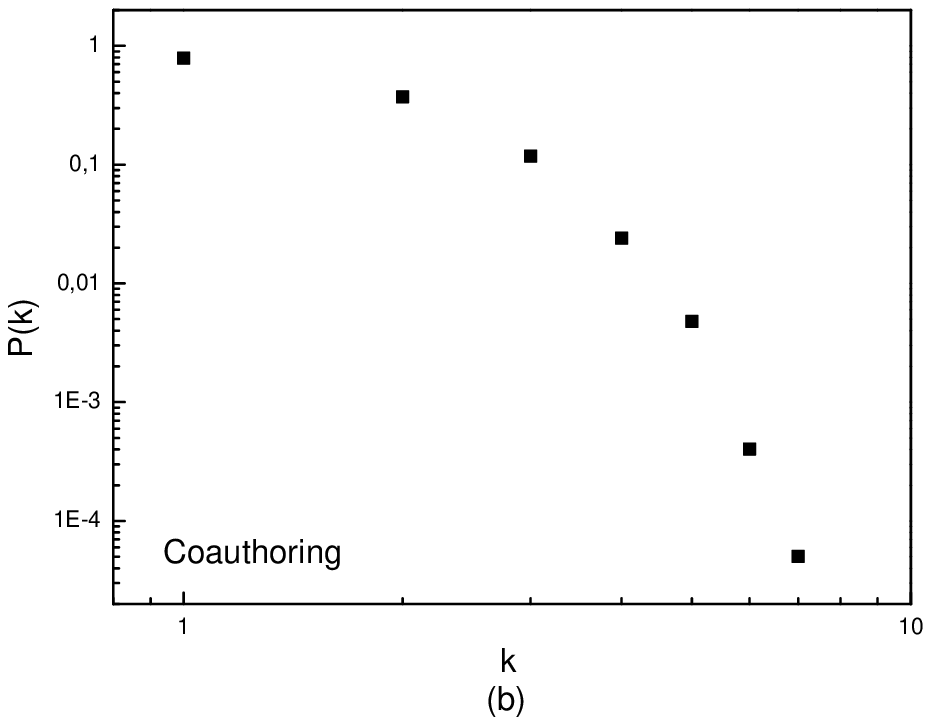}
\caption{(a) Degree distribution of authors in the Coauthoring
network. The degree $k$ corresponds to the number of papers of each
author signed. (b) The same distribution of papers in the
Coauthoring network. The degree $k$ corresponds to the number of
authors of each paper.}\label{Coauthoring}
\end{figure}

The degree distribution of two sets of these four dependence
networks are given in Fig.
\ref{Econophysicists}--\ref{Coauthoring}. Apparently, One may
observe on these plots that all these distributions have a
property in common: the degree distributions of actors (Shown in
Fig. \ref{Econophysicists}(a)--\ref{Coauthoring}(a)) are very
heterogeneous and fit power laws very well in all cases \cite{
basic,self,model,model1}. On the contrary, the degree
distributions of groups (Shown in Fig.
\ref{Econophysicists}(b)--\ref{Coauthoring}(b)) are far from a
power law in these four cases. The reason may be the limit number
of authors, words in papers and sentences.

\subsection{The Independence Networks}

{\bf{Library}} Here we analyze a books-readers bipartite network
which the database obtained from Beijing Normal University library
during $2005$, with $N_{readers}=17593$ and $N_{books}=91752$. An
edge exists between a reader and a book if he/she borrowed this
book, and $E=369394$ in this graph. The average degree of readers
and books respectively are $<k_{readers}>=20.99$ and
$<k_{books}>=4.026$, while the maximal degree are $410$ and $361$.

{\bf{Maze}} It is a peer to peer exchange bipartite network. In
this bipartite network, the data obtained from $Maze.com$ by
registering all the exchanges processed by a large server during
one week. There are two kinds of peers to be two kind nodes. One
kind peers are the downloaders who download the resources which
provided by the other kind peers. It consists of
$N_{downloader}=110163$, $N_{resources}=25171$, and $E=924738$.
The average degree of downloaders and resources respectively are
$<k_{downloaders}>=8.39$ and $<k_{resources}>=36.74$, while the
maximal degree are $1455$ and $1668$.

{\bf{P2P}} It is also a peer to peer exchange bipartite network.
But the differences with $Maze$ are not only the data source but
also the definition of two kind nodes. Here peers and data are two
different kinds of nodes. The data source gotten from all the
exchanges processed by a large server during $48$ hours, contains
$N_{peers}=1986588$, $N_{data}=5380546$, and $E=55829392$. The
average degree are respectively $<k_{peers}>=28.1$ and
$<k_{data}>=10.38$.

\begin{figure}
\includegraphics[width=7cm]{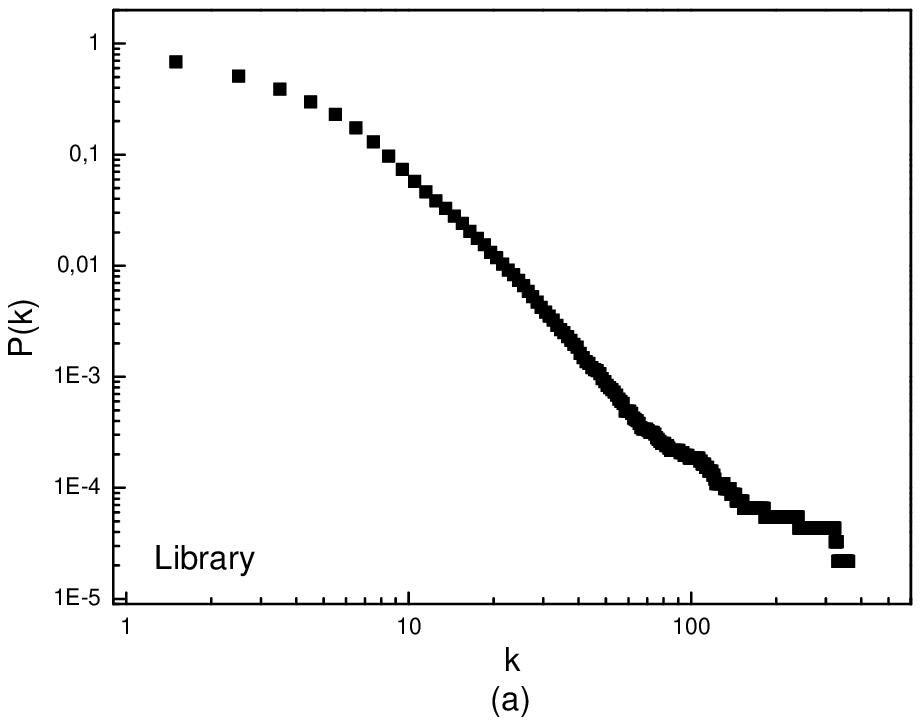}\includegraphics[width=7cm]{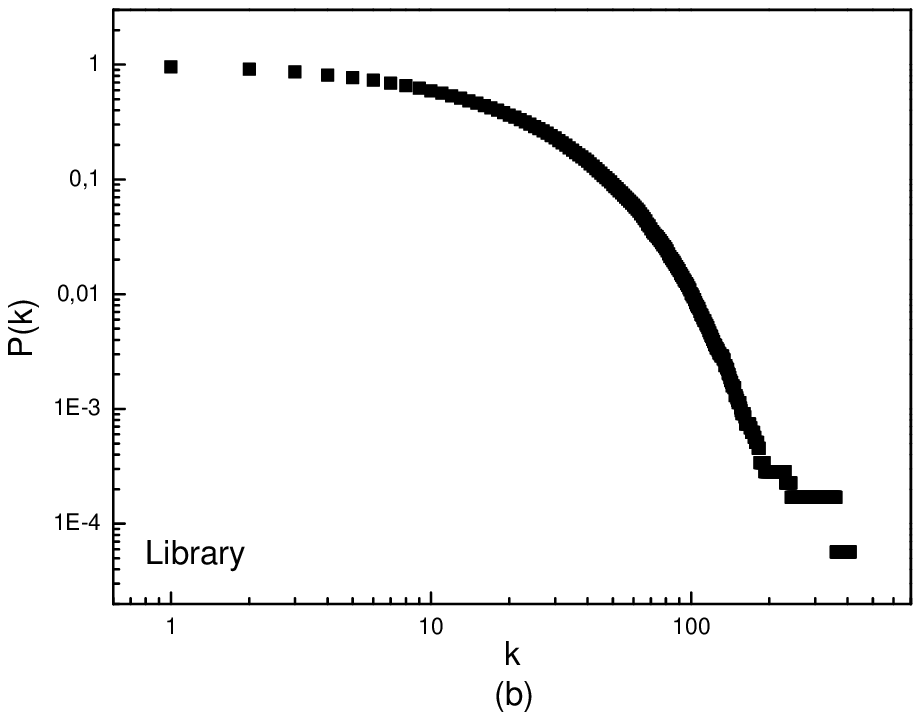}
\caption{(a) Degree distribution of books in the Library network.
The degree $k$ corresponds to the number of readers of each book
have been borrowed. (b) The same distribution of readers in the
Library network. The degree $k$ corresponds to the number of books
of each reader have borrowed.}\label{Library}
\end{figure}

\begin{figure}
\includegraphics[width=7cm]{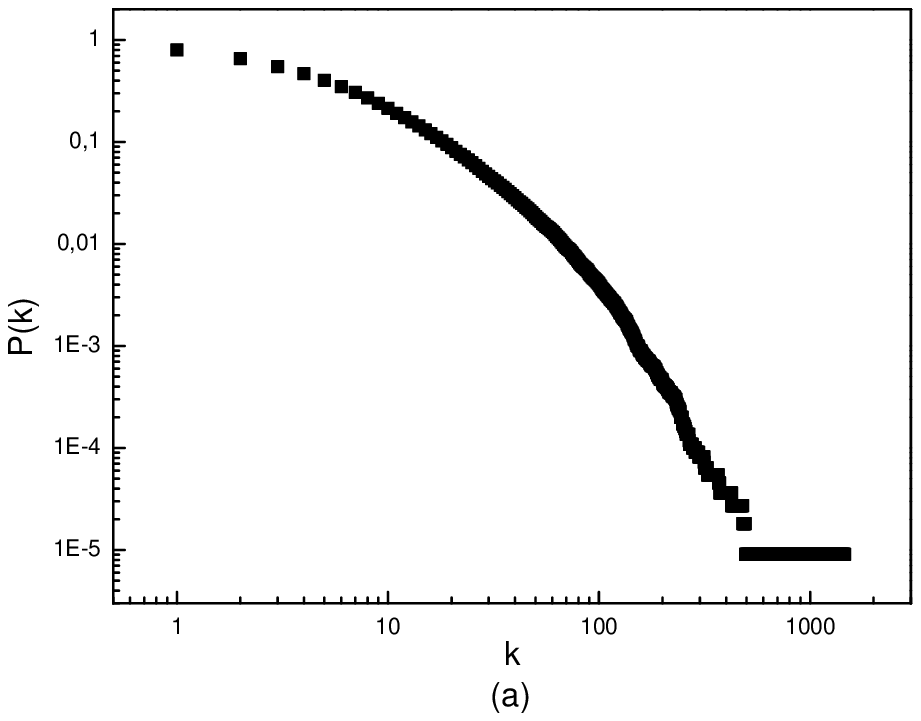}\includegraphics[width=7cm]{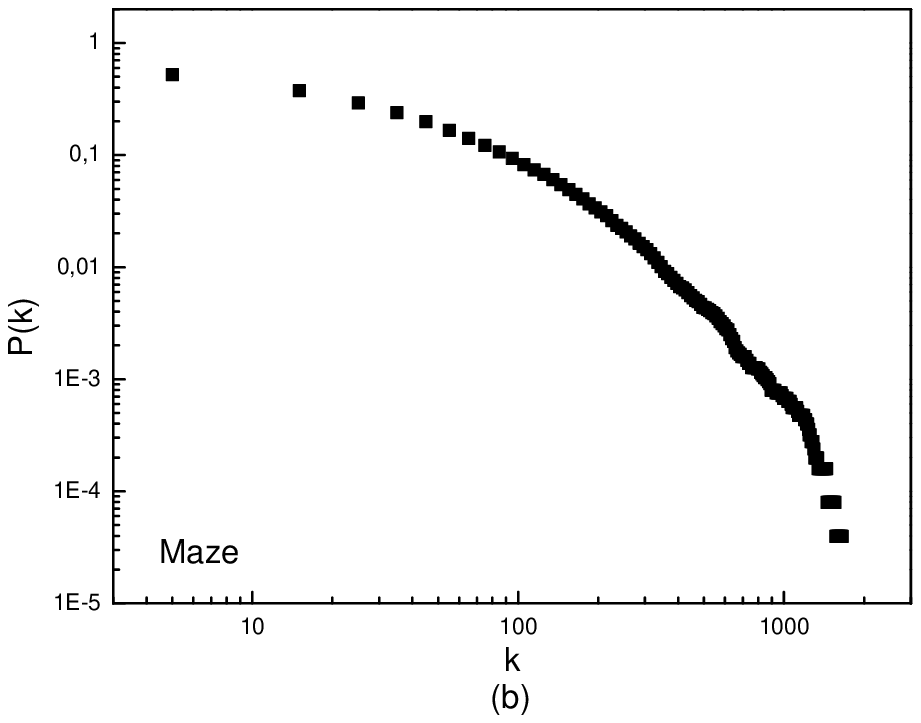}
\caption{(a) Degree distribution of downloaders in the Maze network.
The degree $k$ corresponds to the number of resources of each
downloader have downloaded. (b) The same distribution of uploaders
in the Maze network. The degree $k$ corresponds to the number of
dowloaders of each uploader shared.}\label{Maze}
\end{figure}

The degree distribution of two sets of $Library$ and $Maze$ are
given in Fig. \ref{Library},\ref{Maze}. The real world databases
of independence networks are very limit which we could collect.
Just from the gotten results in these three cases, it hardly
concludes that there are some common properties in the
independence networks.

\section{The Evolution Model and Simulation Results}\label{model}

\subsection{The Evolution Model}

In order to understand the scale-free property of actors in
dependence networks, most of the former works used the
preferential attachment principle to define the models
\cite{paper,self,model,model1,lambiotte}. Here we propose two
growing models without the preferential attachment principle to
obtain that the authors degree distributions generally follow a
power law. To be concrete, we will describe the developments in
terms of authors and papers, but our results are applicable to any
real affiliation networks and generated networks. The first
growing bipartite network model (Model $1$) is defined by the
following rules:

(1) At each time step a new paper with $n$ authors is added.

(2) In these $n$ authors, there is one author is new with
probability $\lambda$, or not in this time step.

(3) The rest $n-1$ or $n$ authors are chosen from the pool of
"old" authors. The rule follows: randomly select an old existing
author, every other authors is randomly selected from its $m$th
nearest neighborhoods.

The first two steps of the second model (Model $2$) are the same
as the first one. The big difference between them is the rule of
selecting the rest authors.

(3) The rest $n-1$ or $n$ authors are chosen from the pool of
"old" authors. The rule follows: randomly select an old existing
author, every other authors is selected by random walk from it by
$m$ steps.

where the parameter $\lambda$ is a constant. The number $n$ is a
random number gotten from $1$ to $max_{n}$. The number $m$ is also
a random number gotten from $1$ to $max_{m}$. $max_{n}$ and
$max_{m}$ are two constants.

\subsection{The Simulation Results}

\begin{figure}
\includegraphics[width=7cm]{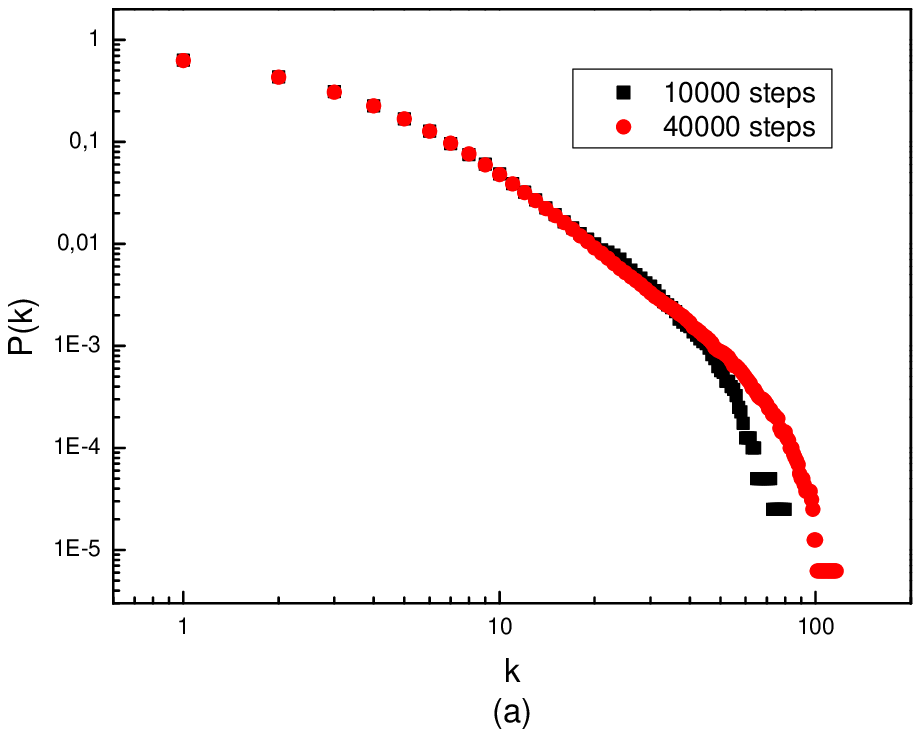}\includegraphics[width=7cm]{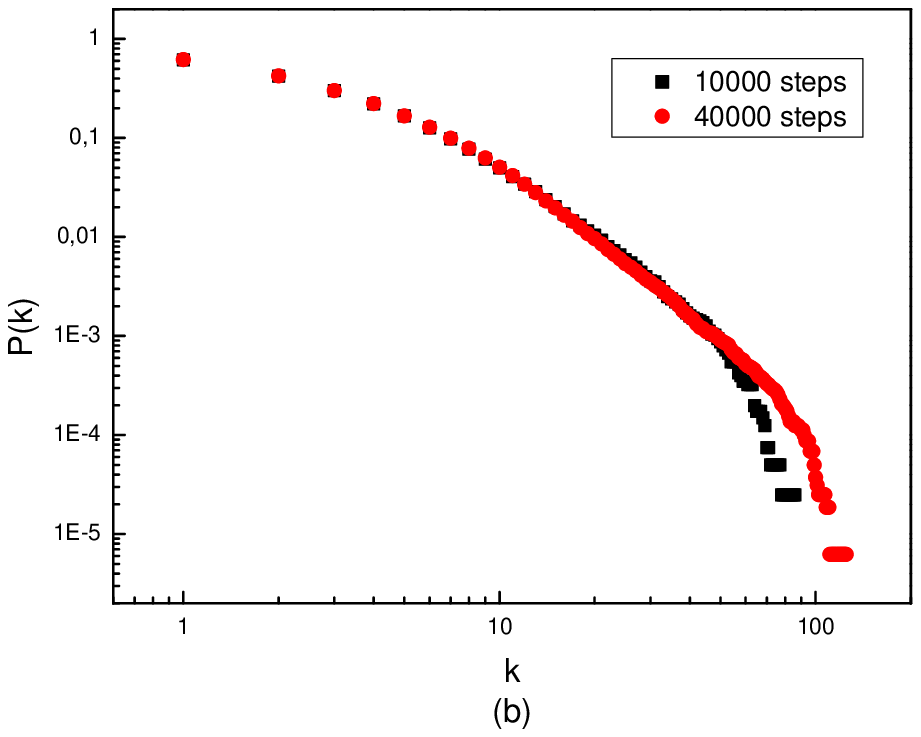}
\caption{(a) Degree distribution of authors in the Model $1$ with
different system size $10000$ steps and $40000$ steps. (b) Degree
distribution of authors in the Model $2$ with different system size
$10000$ steps and $40000$ steps.}\label{step}
\end{figure}

\begin{figure}
\includegraphics[width=7cm]{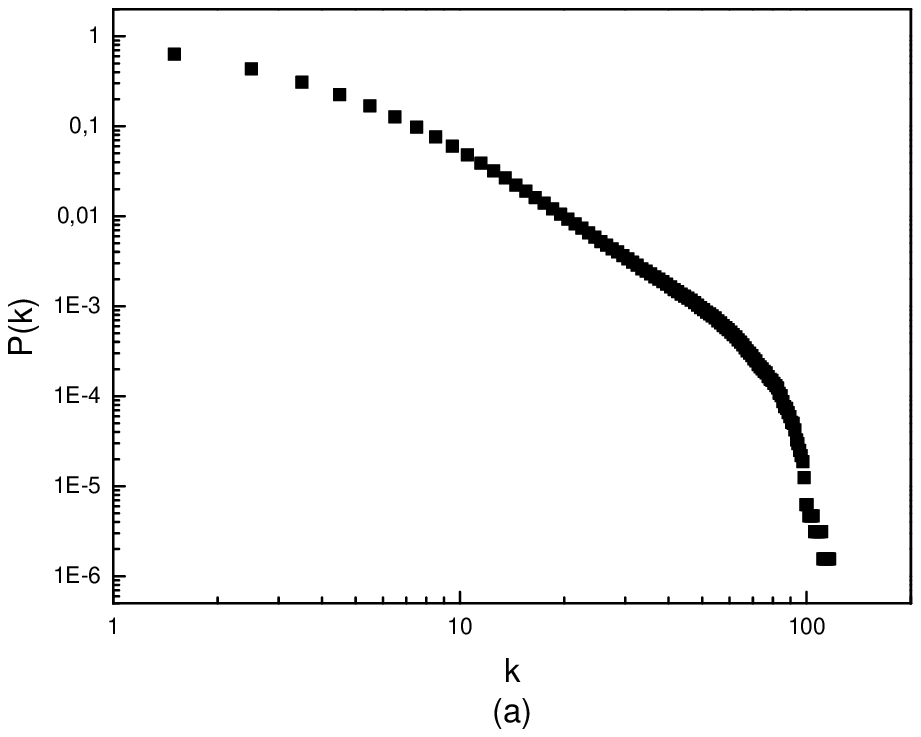}\includegraphics[width=7cm]{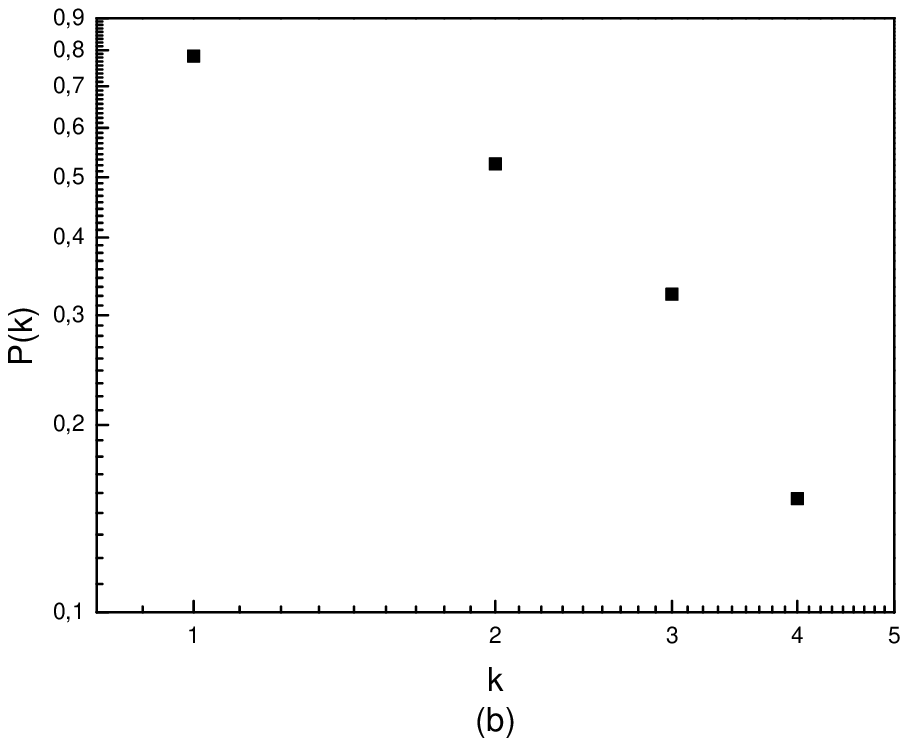}
\caption{(a) Degree distribution of authors gotten by Model $1$ with
parameters $max_{n}$=$5$, $\lambda$=$0.8$, $max_{m}$=$2$. It obeys a
pow law well. (b) Degree distribution of papers gotten by Model $1$
in the same parameters.}\label{model1}
\end{figure}

\begin{figure}
\includegraphics[width=7cm]{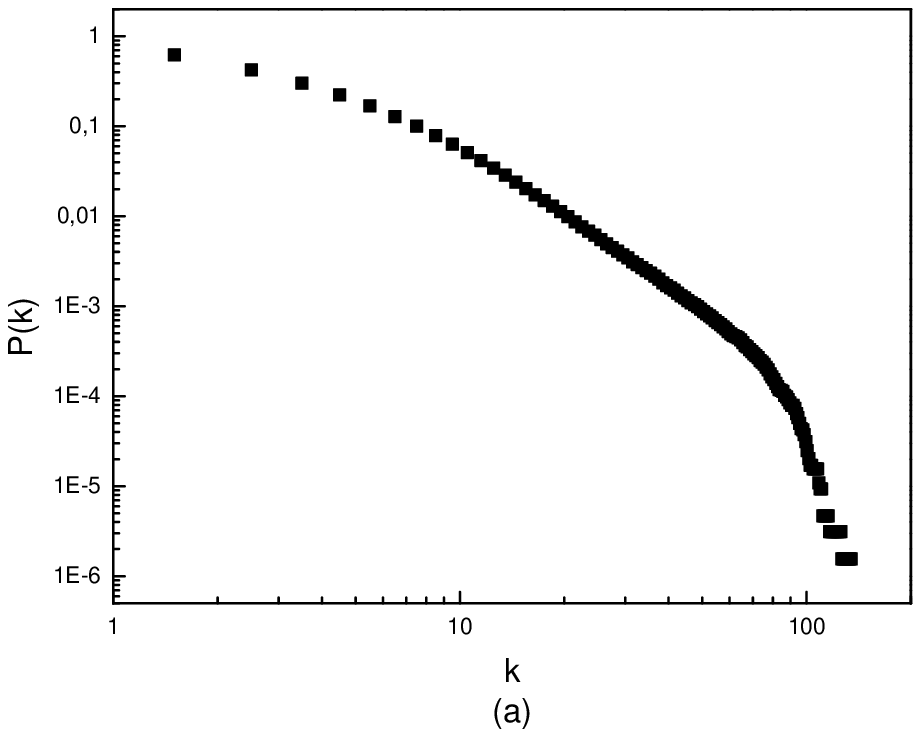}\includegraphics[width=7cm]{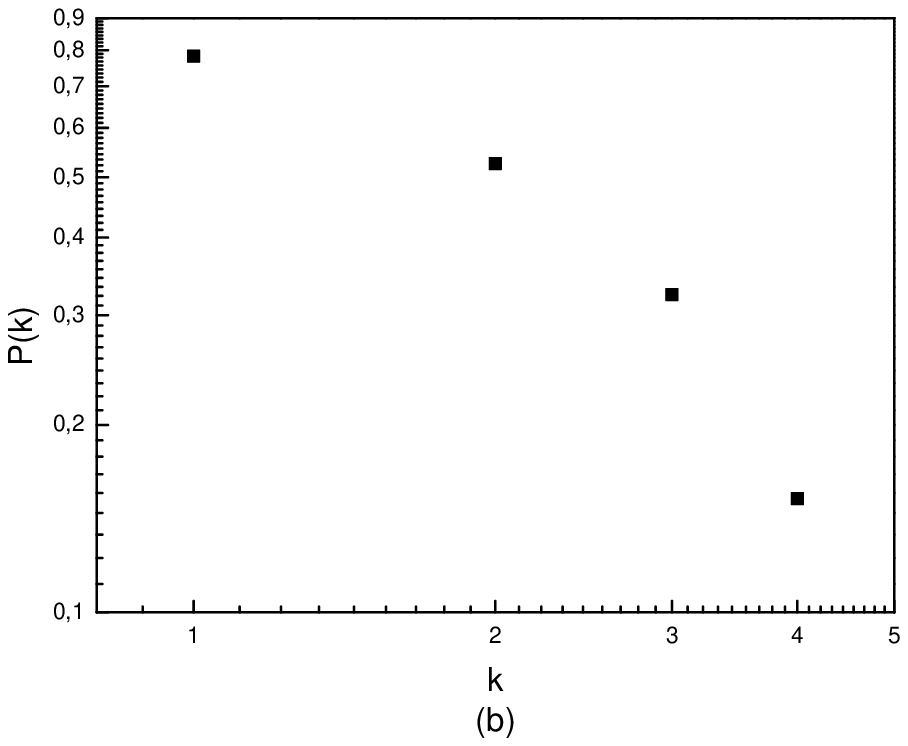}
\caption{(a) Degree distribution of authors gotten by Model $2$ with
parameters $max_{n}$=$5$, $\lambda$=$0.8$, $max_{m}$=$2$. It also
obeys a pow law well. (b) Degree distribution of papers gotten by
Model $2$ in the same parameters.}\label{model2}
\end{figure}

\begin{figure}
\center \includegraphics[width=9cm]{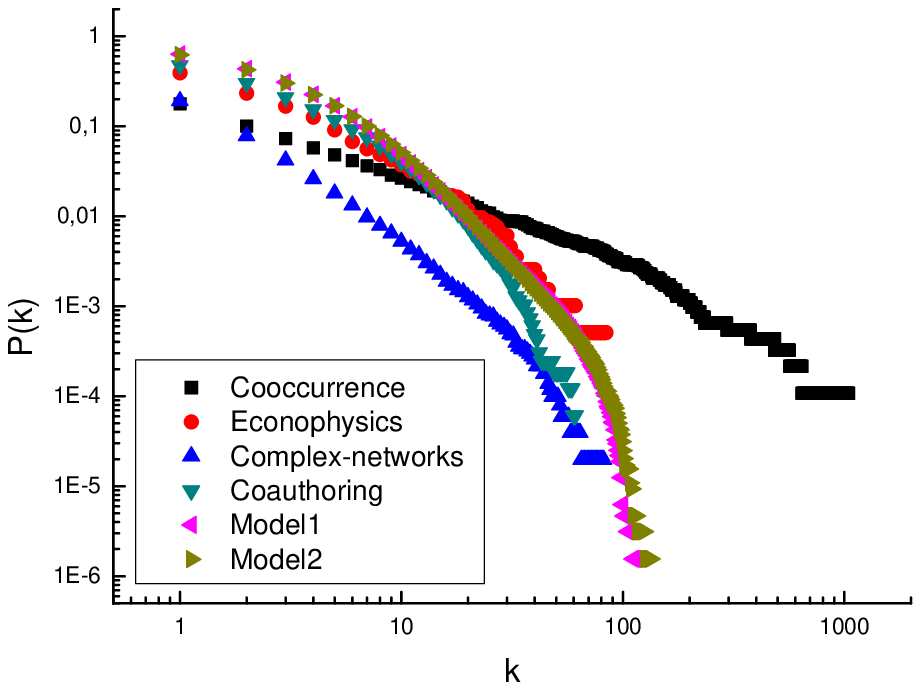} \caption{Degree
distribution of actors, including empirical results from
$Econophysicist$, $Complex-networks$, $Cooccurrence$, $Coauthoring$
and simulation results from $Model 1$, $Model 2$.}\label{compare}
\end{figure}

In this section, The numerical results of Model $1$ and Model $2$
are apart given in Figs. \ref{step}--\ref{compare}. All the
results are the average of $20$ simulations for different
realization of networks under the same parameters. The step of all
generated networks all reach $40000$ times. We have also did the
simulation to $10000$ steps. The final distributions of
corresponding quantities are almost same (Shown in Fig.
\ref{step}). A network with $40000$ papers could give us a nice
description for asymptotic distribution. Parameters are
$max_{n}$=$5$, $\lambda$=$0.8$, $max_{m}$=$2$ for two models if
not mentioned. As the increasing of $max_{n}$, $max_{m}$ and the
decreasing of $\lambda$, the degree distribution of authors is far
from a pow law distribution in theses two models. Fig.
\ref{model1} and \ref{model2} show the cumulative degree
distribution of authors and papers obtained by Model $1$ and Model
$2$ respectively. The simulations are consistent with the scale
free property observed from empirical data. To check the validity
of our two models, we also proceed to compare the empirical data
on $Econophysicist$, $Complex-networks$, $Cooccurrence$,
$Coauthoring$ networks with numerical simulations of two models
made above (see Fig. \ref{compare}).

\section{Conclusion}\label{conclusion}
Many real-world networks display a natural bipartite structure.
This makes us study them should be as bipartite ones. In this
paper, we collect some representation real-world bipartite
networks and according to the different relation of two sets of
nodes, divide them into two types, dependence bipartite networks
and independence bipartite networks. In dependence networks, one
kind of nodes is the results or groups which are gotten by some
kind of behaviors among the other kind of nodes. The independence
network is opposite with the dependence network, two sets exist
independently. By analyzing the degree distribution of these
collected networks, we find the actors nodes have scale-free
property in the dependence networks, and there is no accordant
degree distribution in the independence networks for both two
types of nodes.

Noticing the scale-free property of actors in dependence networks,
we propose two models for generation of dependence bipartite
graphs. This two models are both growing models, but are not based
on the preferential attachment mechanism. The obtained networks
fit well the properties of empirical results, using only their
general bipartite structure. However, there still have other
properties not obtained by the bipartite models, The independence
network is not studied well. There are a lot of works to do in
bipartite networks.

\section{Acknowledgement}
The author P. Zhang thanks Matthieu Latapy for sharing the network
data. This work is partially supported by $985$ Projet and NSFC
under the grant No. $70771011$ and No. $60534080$.

\end{document}